\documentclass{article}
\usepackage{epsfig}
\usepackage{amsmath}
\usepackage{amsfonts}
\usepackage{amssymb}
\usepackage{euscript}
\usepackage{graphicx}
\begin{document}

\sloppy

\sloppy

\begin{center}
{{\Huge\bfseries The new thermo-magnetic effect in metals
}}

\vspace{1cm}

\itshape{ B.V.Vasiliev}\\
bv.vasiliev@narod.ru
\end{center}

\begin{center}
Abstract
\end{center}
The magnetic field which induced by the  thermo-electric current in metals  was detected and measured using of a flux-gate magnetometer.
It is shown that the application of a temperature gradient on a metal rod gives rise to a circulating current therein and induces a magnetic field in the vicinity of its surface.
If a temperature gradient on a metal rod  exists,  the "hot"\  electrons flow from the heated region of a metal  into a colder region and
extrude  "cold"  electrons that form a current in opposite direction.
Since the oppositely directed currents repel each other due to the interaction of magnetic fields, a convective loop of electron current formes inside a metalic sample.
The magnetic field of this convection is directly proportional to the temperature gradient, the metal conductivity and inversely proportional to the temperature squared.

\begin{verse}

\vspace{1cm}

\hspace{3cm} However, the thermopower is not a very valuable\\
\hspace{3cm} probe of fundamental electronic properties\\
\hspace{3cm} of a metal.\\

\vspace{.5cm}
{\it
\hspace{5cm}N.Ashcroft, N.Mermim:\\
\hspace{5cm} Solid State Physics,v.1, p.258.\\
\hspace{5cm}Holt, Reinhart and Winston,(1976)\\}
\end{verse}

\section{Introduction}
The thermoelectric phenomena which occur when gradient of temperature  applied on a conducting specimen together  with an external magnetic field is usually accepted to call as thermomagnetic effects.

There are a number of such  phenomena which are well studied, for example, it is the  Nernst-Ettingshausen effect or the  Righi-Leduc effect.
A number of these effects are discussed in detail in the monograph \cite{Callen}. \\
It is important to stress that all these phenomena are exist  in semiconductors only.\\

The effect described below has a different nature.\\
In fact we will discuss  magnetic fields which are directly induced by gradient of temperature in  metal specimens.
Exactly, one can say, that we will consider magnetic  fields of thermoelectric currents in metals.\\
Thermoelectric effects in metals are well studied \cite{Ashkr}, \cite{Kitt}.\\
Originally the process of the heat transfer in electron gas of metals was described by  Drude  \cite{Ashkr}. Late the Drude`s model has been enhanced by Lorentz and Sommerfeld.

However, the magnetic field produced by a thermoelectric current of  conduction electrons  apparently has never been observed and investigated.
Probably, the reason for this is  that  the courses related to this issue contain a some misunderstanding.
The high thermal conductivity of metals is result of the ability of free electrons to  transport a heat.
 At this process, the "hot"\  electrons flow from the heated region of a metal  into a colder region and
extrude  "cold"  electrons that form a current in opposite direction.
Until now it has been commonly thought that in the case of a homogeneous metal,  both these opposite electron current have a diffusive character and flow uniformly throughout the cross section of the sample.
A deviation from uniformity of  diffusive  flux can be due to inhomogeneities in the sample only.
However, this consideration has a mistake.
Electric currents repel each other due to their magnetic interaction, if they flow  in opposite directions.
As a result, a peculiar  electron convection occurs inside a metalic sample (see Fig.\ref{C1}), which leads to the existence of the magnetic field in the vicinity of the sample.
The researching of this field is a focus of this work .

Experimentally this magnetic field is quite large and it can be detected by conventional flux-gate magnetometer.
Therefore, when measuring small magnetic effects in metals with using SQUID-magnetometers, experimenters must  to keep in mind that the specifying thermomagnetic effect can influence under certain conditions on the measurement results.

\section{Thermoelectric currents in a metal.}

The system of  conducting electrons in a metal has the total density:
\begin{equation}
n_e = z n_i
\end{equation}
 where $ z $ is the number of conducting electrons on a metal ion (the valence), \\ $ n_i $ is the ion density in a metal.

 As it was noted by A.Sommerfeld, all conducting electrons of a metal are divided into two groups. \\

 In equilibrium state at temperature T, there are  non-degenerated electrons   with density $n_a$, which are activated by a heat. 
 They occupy levels above the Fermi level in the energy distribution.

The second group is formed by degenerate conducting electrons that occupy the energy levels below the Fermi level and have a density:
 \begin{equation}
n_c = n_e-n_a.
\end{equation}

 Thermally activated non-degenerate electrons have the density:
 \begin{equation}
n_a\approx n_e \frac{T}{T_F}.
\end{equation}
They have the ability to descend from one energy level to another, and so they can  transfer the heat at their moving   in a metal.

Due to the fact that we consider a metal at room temperature ($T\ll T_F$), the group of degenerate electrons  includes almost all conducting electrons $ n_c\cong n_e$.
 These electrons can not  be involved in a process of a heat transfer.
But, if the electric field exists inside the metal, they create a current.
And they determine the electrical conductivity of a metal.

If the temperature gradient $\mathbf{\nabla}T$ is attached to a metal rod, it will  activate a thermal current of electrons from a hot region, where the electrons have higher speed and higher pressure.

Thermally activated electron  betrays a transferring heat to the lattice due to inelastic collisions with phonons. Therefore, the free path of hot electrons between two successive collisions with phonons is inversely proportional to the phonon density $n_{ph}$:
\begin{equation}
\frac{1}{l_{ph}}\approx n_{ph} \cdot S
\end{equation}
Where $S$ is the electron-phonon scattering cross section.
Due to the fact that we consider the metal at room temperature, and given the fact that the Debye temperature in most metals is of the same order of value, we can assume that
\begin{equation}
S\approx n_i^{-2/3}\label{S}.
\end{equation}

The energy of the phonon gas in the Debye approximation \cite{Kitt}
\begin{equation}
\mathcal{E}_{ph}= \frac{3\pi^4}{5}n_i k T_D \left(\frac{T}{T_D}\right)^{4},\label{eph}
\end{equation}
where $T_D$ is the Debye temperature.

The phonon energy is approximately equal to $kT$, so
  their density can be described by:
\begin{equation}
n_{ph}\approx \frac{3\pi^4}{5}n_i \left(\frac{T}{T_D}\right)^{3},
\end{equation}

Subject to Eq.(\ref{S}) phonons limit the mean free path of "hot" electrons by the value:
\begin{equation}
{l_{ph}}\approx \frac{5}{3\pi^4}n_{i}^{-1/3} \cdot\left(\frac{T_D}{T}\right)^{3} 
\end{equation}

As the  velocity of conducting electron  is approximately equal to the Fermi velocity, the free path of electrons between two successive collisions with phonons is:
\begin{equation}
\tau_{ph}= \frac{l_{ph}}{v_F}\approx  \frac{\frac{5}{3\pi^4}}{n_i^{1/3} v_F} \cdot\left(\frac{T_D}{T}\right)^{3}\label{taup}.
\end{equation}

To calculate the average velocity of the electrons we assume that  two non-degenerate electrons coming from  hot end and cold ends of a rod  are thermalized in  the point $x$ inside this metal rod.
If the difference in their velocities are small, their average speed in the one-dimensional model can be written as:
\begin{equation}
\it{v}_+=\frac{1}{2}\left[\it{v}(x-\it{v}\tau_{ph})-\it{v}(x+\it{v}\tau_{ph})\right]=-\tau_{ph} v\frac{dv}{dx}=-\tau_{ph}\frac{d}{dx}\left(\frac{v^2}{2}\right).
\end{equation}
Turning to the case of three dimensions, we can write \cite{Ashkr}:
\begin{equation}
\mathbf{v}_+=-\frac{\tau_{e}}{6}\frac{dv^2}{dT}\nabla T=-\frac{\tau_{e}}{6}\frac{c_v}{m_e n_e}\nabla T,
\end{equation}
where
\begin{equation}
c_v=\frac{\pi^2}{3}kn_{e}\frac{kT}{\mathcal{E}_F}
\end{equation}
is the Sommerfeld's heat capacity of electron gas,\\
$m_e$ is the electron mass.

Given that electrons are passed to the point of  the energy exchange from the  distances which are equal to the mean free path,  the velocity of thermal diffusion averaged over the whole electron gas becomes:
\begin{equation}
<\mathbf{v}_+>=-\tau_{ph}\frac{\pi^2}{18}\frac{k}{m_e}\frac{T}{T_F}\nabla T.
\end{equation}
The considered heat flow induces the electron current
\begin{equation}
\mathbf{j}_+=en_e<\mathbf{v}_+>\approx\tau_{ph}\frac{\pi^2}{18}\frac{ke}{m_e}n_e\left(\frac{T}{T_F}\right)\nabla T,
\end{equation}
and creates inside the metal rod  the Seebeck's electric field:
\begin{equation}
\mathbf{E}_S = Q_S\nabla T,
\end{equation}
under the action of which there occurs a reverse flow of electrons:
\begin{equation}
\mathbf{j}_-=e n_e v_-=\sigma Q_S\mathbf{\nabla}T,
\end{equation}
where $\sigma$ is the conductivity of a metal.

The average velocity of electrons in this current \cite{Ashkr}:
\begin{equation}
<\mathbf{v_-}> = - \tau_e\frac{e}{m_e}\mathbf{E}_S,
\end{equation}
where $\tau_e$  is the free time of electrons moving in an electric field.
 
 The temperature dependence of this time is calculated in a number of courses on the theory of metals (see eg \cite{LL}) and it is approximately described by the equation:
 \begin{equation}
\tau_e\approx\frac{\hbar}{kT}.\label{taue}
\end{equation}
Hence we obtain
\begin{equation}
\mathbf{j}_-\approx n_e \frac{\hbar}{kT}\frac{e^2}{m_e}Q_S\mathbf{\nabla}T.
\end{equation}
and taking into account that $\mathbf{j}_+ + \mathbf{j} _-=0 $, we obtain the expression for the Seebeck coefficient:
\begin{equation}
Q_S\simeq -\frac{\pi^2}{18}\frac{k}{e}\frac{T_D^3}{T T_F^2}.
\end{equation}

\section{The testing by means of the Wiedemann-Franz law.}
 The Wiedemann-Franz law establishes the relationship between the electrical conductivity of the metal and its thermal conductivity.
The compliance of obtained estimations  with this law  should indicate on their correctness.  Let us check this criterion.

The thermal conductivity of gas is determined by the heat capacity $ C $ of the environment that takes the heat from the hot gas particles, the gas particle velocity $ v $ and the length of its free path $ l $ \cite{Kitt}:
\begin{equation}
\kappa=\frac{1}{3}C v l\approx \frac{1}{3}C v^2 \tau.\label{kap}
\end{equation}

In the case of atomic or molecular gas the thermal conductivity is determined by their specific heat.
The case of an electron gas is more complicated.
 The electron-electron interaction  is weak in metals and the main mechanism of heat transfer is the electron-phonon interaction.
The "hot" electrons transfer their energy to phonons. The phonon gas is the medium that takes the energy from the electrons.  So in this case  the specific heat of the phonon gas determines the electron gas thermal conductivity, which according to Eq.(\ref{eph}) approximately described by the equation:
\begin{equation}
C_{ph}\approx \frac{12\pi^4}{5}n_i \left(\frac{T}{T_D}\right)^{3},
\end{equation}
In view  of the mean free time of  non-degenerate electrons Eq.(\ref{taup}), we get:
\begin{equation}
\kappa=\frac{2}{3}k n_i^{2/3} v_F.\label{kap1}
\end{equation}
Taking into account Eq.(\ref{taue}), we have:
\begin{equation}
\sigma T = \frac{e^2n_e}{m_e}\frac{\hbar}{k}\label{sigm1}
\end{equation}
and thus we obtain the ratio:
\begin{equation}
\frac{\kappa}{\sigma T}\approx 4\frac{k^2}{e^2},
\end{equation}
which is in good agreement with the Wiedemann-Franz law.

\section{The electron convection in metals.}

Therefore, the gradient of the heat in the metallic rod with the conductivity $\sigma$ induces the current density:
\begin{equation}
\mathbf{j}_-=-\mathbf{j}_+=\sigma \mathbf{E}_S \simeq \sigma\frac{T_D^3}{T T_F^2}\frac{k}{e}\nabla T.\label{si}
\end{equation}

\begin{figure}
\hspace{1cm}
\includegraphics[scale=0.5]{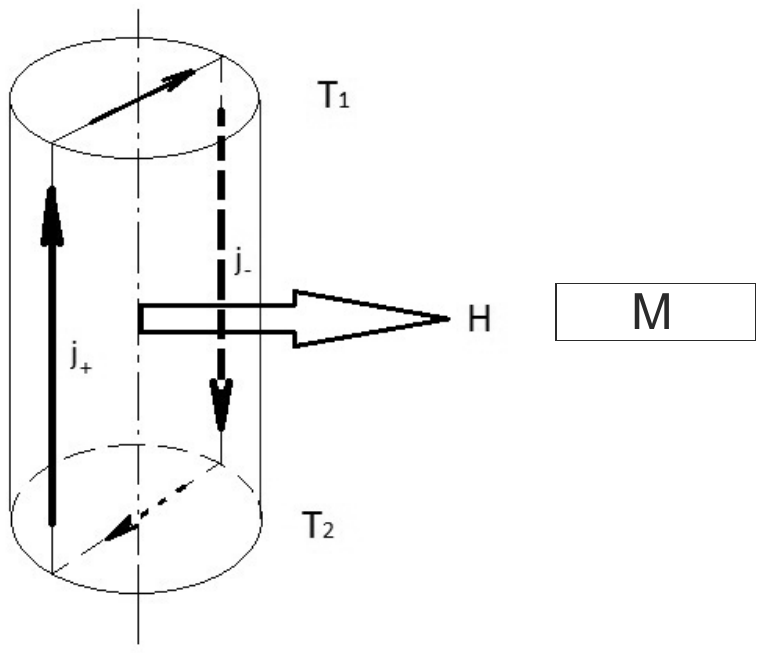}
\caption {Currents induced by the temperature gradient in cylindrical metallic sample; M is the magnetometer.}\label{C1}
\end{figure}

The magnetic interaction repels the electric currents if they flow in opposite directions.
 Therefore, in the case of a cylindrical specimen, the ends of which have different temperatures, currents $j_+$  and $j_-$ must flow through the diametrically opposite sides of cylinder (Fig.\ref{C1}). \footnote{This distribution of currents inside the metal body in its form is similar to the convective flow in the gas. It allows to call   this phenomenon as "convection" for the sake of brevity, although the physics of these phenomena are quite different of course.}

Currents which was created by the gradient of the heat induce   near the outer surface of the cylinder  the magnetic field with intensity:
\begin{equation}
H\approx\frac{2j\cdot \pi R^2}{cR}\approx \Theta_R\left(\frac{\nabla T}{T^2}\right)\label{H},
\end{equation}
where the constant
\begin{equation}
\Theta_R\approx \frac{\pi^3 }{9 }\frac{k}{ce}(\sigma T)\frac{T_D^3}{T_F^2}R\label{ce},
\end{equation}
where $R$ is the radius of cylinder.

\section{The measurement results}
When the temperature gradient of about 1 $\frac{grad}{cm}$, which is not too difficult to create on a copper sample at room temperature, the induced magnetic field near the sample surface may reach $10^{-3} Oe$, and it is not necessary to use a highly sensitive SQUID-magnetometer for its observation with which this effect was first observed by us.

 The fluxgate-magnetometer is more simple, is more suitable for measurements in room temperature range and has quite enough sensitivity.

 In our experiments \cite{V-G}, the fluxgate-magnetometer was placed near the middle of the metallic cylinder height. The temperature gradient was applied to cylinder. The temperatures of its ends was automatically measured at frequent intervals to determine the temperature gradient and the average temperature of the cylinder. The cylinder can rotate around its axis and fixed in a selected position for the registration component of the magnetic field perpendicular to the cylinder axis (see Fig.\ref{C1}).

\subsection{The determining of the "convective loop of current" \ orientation}
\begin{figure}
\hspace{1cm}
\includegraphics[scale=0.5]{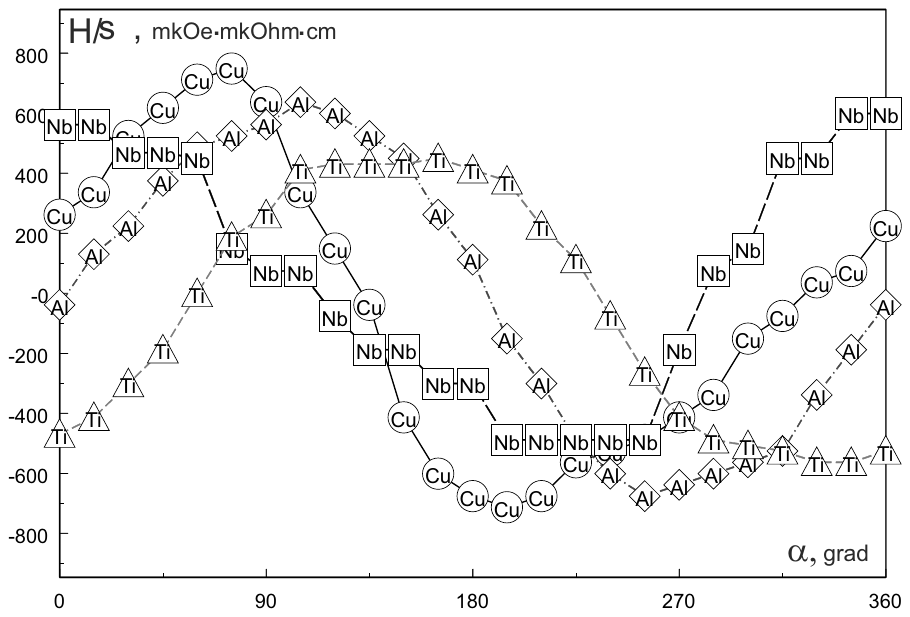}
\caption {The angular dependence of the induced "convective" \ magnetic field, referred to the conductivity of the metal. The shift in the angle for different samples in this figure due to the fact that in each case, the measurement started with an arbitrary angular orientation of the cylinder. The angle of turnabout of the sample is shown on the horizontal axis. The induced magnetic field, referred to the conductivity of the metal, is shown on the ordinate.}\label{fig3g}
\end{figure}

The first step in the investigation of the phenomenon of convection in the electron gas was the   determination of the "loop of current" \ orientation  inside the cylindrical sample. To do this, the cylinder has consistently turned a small angle (approximately equal to $15^o$) and measuring the projection of the induced field on the axis magnetometer was made. The measurement results are shown in Fig. \ref{fig3g}. As it can be seen, the angular dependence of the induced field has the sinusoidal character.
The shift in the angle for different samples in this figure due to the fact that in each case, the measurement started with an arbitrary angular orientation of the cylinder.\\

After these measurements, the  marks were deposited on the surface of the cylinder  to denote the "loop of current" \ orientation. \\

Subsequent measurements were made at the particular orientation of the "loop of current" \ , at which the  maximum of the field  was obtained. \\

It should be emphasized that the position of "the convectional loop of current" \ inside the cylinder during the entire measurement period remained unchanged. \\

Special efforts to change the orientation of the "loop of current" \  have been made: \\

a) the significant hardening at the end face of the copper cylinder was made with a hammer, \\

b) a step at the end face of the cylinder was manufactured on the milling machine, \\

c) the copper cylinder was first heated to about $500^o$ C and then cooled either slowly or by quenching in water. \\

There was no any visible  effect of  these procedures
on the "loop of current" \ orientation and the reason why "the convectional loop of current" \ has a continuing orientation in the cylinder remained unclear.

It should be noted, that in one of the cooper cylinder,  "the convectional loop of current" \ was not originally found. But the normal effect appeared  after a few light taps with a hammer on the end face of the cylinder.

\subsection{The  "convection" \ of electron gas  in different metals}
The thermo-magnetic effect was  measured on  four cylinders with the diameter of 30 mm and the length of about 150 mm from copper, duralumin, niobium and titanium.
The linear dependence of the "convective" \ magnetic field from the applied temperature gradient is observed for all these metals (Fig. {\ref{naklong}).

\vspace{0.5cm}Table (1).\\
{The thermo-magnetic effect in long cylinders with $R=1.5$ cm at $T\approx 300K$.}

\begin{tabular}{||c|c|c|c|c|c||} \hline\hline
&&&&&\\
metal& $T_F,10^{4}$K & $T_D$,K  &  $\Theta_{calc} $  &  $\Theta_{meas}=\frac{H \cdot T^2}{\nabla T}$   &   $\frac{\Theta_{calc}}{\Theta_{meas}}$\\
 &                         &&Eq.(\ref{ce})& &\\\hline
  Cu  &    8.1   &     343 & 48 & 43.5&1.1 \\
  Al  &   13.4   &     428 & 21 & 9.8 &2.1 \\
  Ti  &   12.9   &     420 & 3.6 & 1.89 &1.9 \\
  Nb  &    9.7   &     275 & 2.0 & 1.38& 1.5 \\\hline\hline
\end{tabular}
\label{HH}

\begin{figure}
\hspace{1cm}
\includegraphics[scale=0.3]{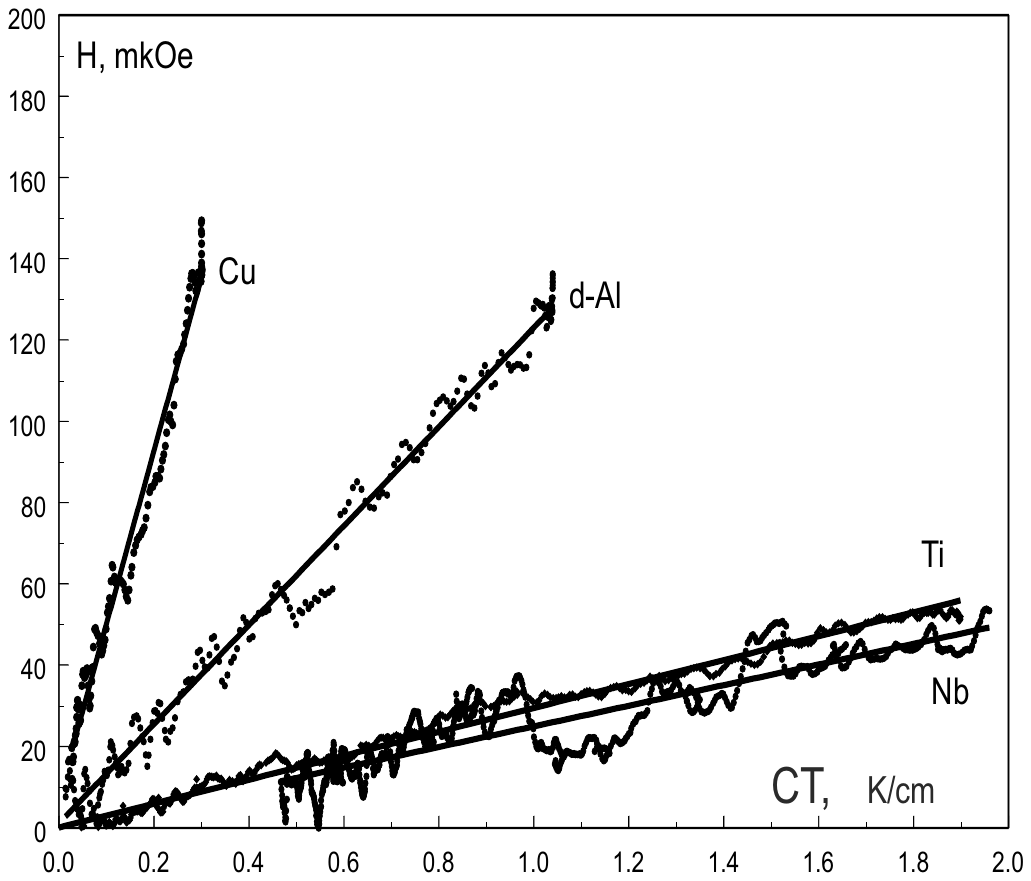}
\caption{The dependence of the thermo-magnetic effect from the applied temperature gradient. On the abscissa the applied temperature gradient is shown in degrees per centimeter. On ordinate the induced magnetic field is plotted in the micro-gauss. The direct line carried out by least squares method.}\label{naklong}
\end{figure}

\newpage

These measurements showed that the above estimate of the magnetic field  induced by the heat flux  is quite satisfactory agreement with measured data (see Table(1)).

\subsection{The temperature dependence of induced magnetic field}
The inverse quadratic dependence of the "convective" \ magnetic field which was predicted by  Eq.(\ref{H}) was tested on the copper sample.
As it can be seen from Fig.\ref{Cu-T2}, the agreement  of the observed data  with the theoretical estimation can be considered satisfactory. \\
\begin{figure}
\hspace{1cm}
\includegraphics[scale=0.5]{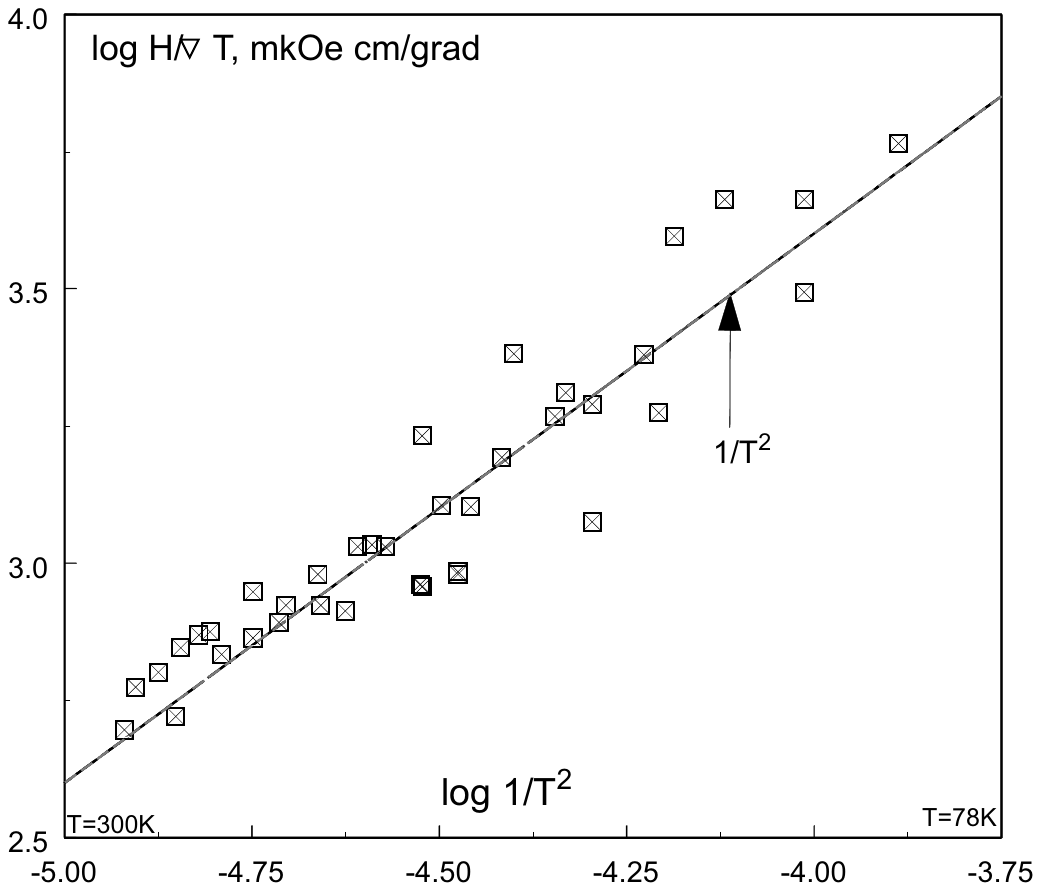}
\caption {The temperature dependence of the "convective" \ magnetic field (referred to the temperature gradient) induced by the copper cylinder.
On abscissa the logarithm of $\frac{1}{T^2}$ is plotted.
On ordinate the logarithm of the magnetic field per temperature gradient in the $mkOe\cdot cm/grad$.
The line shows the dependence  $T^{-2}$.}\label{Cu-T2}
\end{figure}

\section{Conclusion}
The above described thermo-magnetic effect is significant for good conductors.
It can interfere with the study of other phenomena, where small magnetic effects in conductors are measured at the presence of temperature gradients.
It can be assumed that this effect was observed previously in a number of measurements made with the help of high sensitive SQUID-magnetometer.

For example, in \cite{V}, we measured the small magnetic field resulting from rotation of the metal sample and it was proportional to its speed of rotation. It was supposed that this small field could be an effect of the inertial forces of rotation.  However, it might seem that it was the result of excessive friction in one of the cryogenic bearings. The friction in one of bearing could create a temperature gradient on the metal sample and could induce "the loop of current"\ .

In \cite{Zav}, SQUID-magnetometer measured the magnetic field which was produced by a twisting of the metallic crystals to which the temperature gradient was applied. It was evident from the frequent conversation with the author of this work that it can not be excluded that at torsion there was a small  bending of the sample. This bending was altering the projection of "the loop of current" \ on the axis of the SQUID-magnetometer, but this effect was interpreted as the field of twisting. \\

As for the reasons for fixing of the "convective loop of current" \ inside a cylindrical sample, it must be marked that the most simple "geometric" \  reason of it remains unexplored.
The "convective loop of current"\ is formed because the "hot" and "cool" currents repel each other due to their magnetic interaction.
Cylindrical specimens tested in the experiments described above, were made on lathes.
Because of this,  our samples can have a some small ellipticity.
Due to the presence of such ellipticity, it would be energetically favorable for currents  to pave their paths along lines through the ends of the major axis of the ellipse.
It seems that this assumption puts everything in its place, also  it has not been tested in the experiments described above.

The author is grateful A.P.Gavrish for help with the measurements.

\end{document}